\documentclass[fleqn,10pt]{wlscirep}
\usepackage[utf8]{inputenc}
\usepackage[T1]{fontenc}
\usepackage{bm}
\title{Observational constraints on stellar feedback in dwarf galaxies}

\author[1,*]{Michelle L. M. Collins}
\author[1]{Justin I. Read}
\affil[1]{University of Surrey, Physics Department, Guildford, GU2 7XH, UK}

\affil[*]{e-mail: m.collins@surrey.ac.uk}

\begin{abstract}
Feedback to the interstellar medium (ISM) from ionising radiation, stellar winds and supernovae is central to regulating star formation in galaxies. Due to their low mass ($M_{*} < 10^{9}$\,M$_\odot$), dwarf galaxies are particularly susceptible to such processes, making them ideal sites to study the detailed physics of feedback. In this perspective, we summarise the latest observational evidences for feedback from star forming regions and how this drives the formation of `superbubbles' and galaxy-wide winds. We discuss the important role of external ionising radiation -- `reionisation' -- for the smallest galaxies. And, we discuss the observational evidences that this feedback directly impacts galaxy properties such as their star formation histories, metal content, colours, sizes, morphologies and even their inner dark matter densities. We conclude with a look to the future, summarising the key questions that remain unanswered and listing some of the outstanding challenges for galaxy formation theories.
\end{abstract}
\begin{document}

\flushbottom
\maketitle

\thispagestyle{empty}

%\noindent \textbf{Key points:} Please suggest $\sim 5$ key points, which should be single-sentence bullet points that summarize the article and remind readers of the take-home messages. An example of key points can be found at \url{https://www.nature.com/articles/s42254-018-0001-7#Abs3}

%\noindent \textbf{Website summary:} Please suggest an $\sim 40$ word summary for the website.Please begin with a general sentence setting the background, then outline the topics discussed in the article. You can find example summaries at \url{https://www.nature.com/natrevphys/reviews}

\section{Introduction \& context}

Stars are known to directly impact their surrounding interstellar medium (ISM) through the emission of ionising radiation, stellar winds and, at the end of the lives of the most massive stars, through supernovae (SN) explosions \citep[e.g.][]{Tacconi20}. Similarly, central supermassive black holes (SMBHs) can also influence their host galaxies through radiation, winds and jets emitted from their surrounding gaseous accretion discs (known as Active Galactic Nuclei; AGN -- e.g. \cite{Fabian12}). Finally, ionising photons from the first stars and galaxies in the Universe can impact the smallest galaxies, inhibiting or shutting down star formation \citep[e.g.][]{efstathiou92,Wise19}. Collectively, these processes are referred to as `feedback'. Usually, such feedback is assumed to be negative because most of these processes are self-limiting. For example, stellar feedback and external ionising radiation typically heat the ISM reducing further star formation \cite[e.g.][]{Agertz13}, while AGN feedback shuts off the flow of cold gas onto the SMBH preventing further feedback \cite[e.g.][]{sijacki07}. However, positive feedback is also possible, for example when stellar winds compress the surrounding ISM exciting further star formation \citep[e.g.][]{Luisi21}, or when a galaxy-wide wind impacts a hot surrounding corona triggering gas cooling and further star formation \citep[e.g.][]{Fraternali13,Hobbs15,Fraternali17,Vulcani18}. Nonetheless, negative feedback must dominate over positive feedback, at least when averaged over the scale and lifetime of galaxies. Otherwise, star formation would be far more efficient than is observed. Averaged over the whole Universe, galaxies convert just $\sim 6$\% of the available hydrogen into stars, with Milky Way-mass galaxies ($M_* \sim 10^{11}$\,M$_\odot$) being the most efficient and dwarfs ($M_* < 10^{9}$\,M$_\odot$) the least \citep[e.g][]{Read05}. This decreasing star formation efficiency with decreasing galaxy mass already hints at the important role of stellar feedback in regulating galaxy growth.

The importance of stellar feedback for sculpting the properties of galaxies has been well-established for over fifty years. Motivated by the lack of interstellar medium in massive ellipticals, early work by e.g. \cite{matthews71} posited that galactic scale winds driven by SN could explain the absence of ionised and neutral hydrogen observed in these systems. Subsequent works extended this to explain the properties of dwarf galaxies, where feedback could not only account for a lack of inter-stellar medium (ISM), but also the low metal yields and diffuse nature of these systems \cite{larson74,dekel86}. As models and simulations have improved, it is clear that SN and stellar feedback have a substantial impact on not only the luminous properties of the smallest galaxies, but also on their dark matter halos \cite[e.g.][]{navarro96,Read05DMheats,governato10,pontzen12,read16}. Indeed, there is mounting dynamical evidence for this in dwarf galaxies at both low and high redshift \citep[e.g.][]{Read19,Bouche21,Sharma21}. Finally, for the very smallest galaxies ($M_* < 10^{5-6}$\,M$_\odot$), ionising radiation from the first stars and galaxies in the early Universe ($z \sim 8-12$) heats their ISM shutting off further accretion of cold gas \citep[e.g.][]{efstathiou92}. Starved of fresh fuel for star formation, these galaxies then cease to form stars after redshift $z\sim 4$, leaving behind `ultra-faint' dwarfs with only old-age stellar populations \citep[e.g.][]{brown14}. From here on, we will refer to the cessation of star formation -- by any mechanism -- as `quenching'.

In this {\it perspective}, we present direct observational evidences for negative stellar feedback in dwarf galaxies -- defined here to be galaxies with a stellar mass $M_* < 10^9$\,M$_\odot$. We first show how stellar feedback acting on 10-100\,pc scales in individual star forming regions leads to the formation of `superbubbles' and galaxy-wide winds (\S\ref{sec:obs}). We then discuss the impact of stellar feedback on the star formation histories, sizes, colours, metal content and the inner dark matter density of dwarfs (\S\ref{sec:impact}), providing further indirect evidence for feedback while simultaneously highlighting its important role as a key regulator of galaxy formation and evolution. Finally, we conclude with a look to the future and a discussion of what key questions remain unanswered (\S\ref{sec:future}).

\section{Direct evidence for stellar feedback in dwarf galaxies}
\label{sec:obs}

\begin{figure}[ht]
\centering
\includegraphics[width=\linewidth]{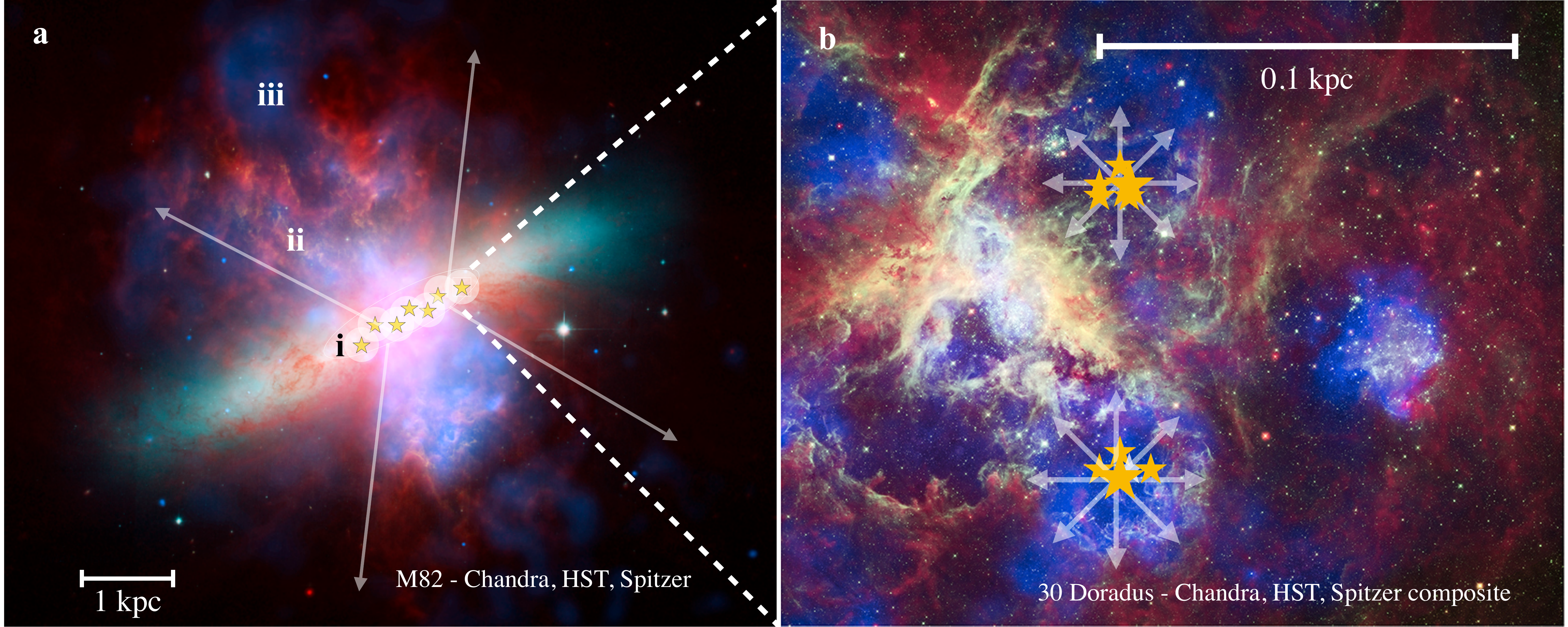}
\caption{{\bf Images of the M82 galaxy and the 30 Doradus star forming region in the LMC.} Both combine Chandra, Spitzer and HST imaging,
and allow us to see how stellar feedback – that originates in star forming regions – impacts galaxies at all scales. 
(a) The M82 galaxy showing a prominent galaxy-wide wind [44]. In an actively star forming (or starburst) galaxy,
bubbles form across multiple sites, overlapping to form larger ‘superbubbles’ (region i). Superbubbles can overlap to form a
large-scale galactic wind (region ii). Galactic winds remove gas from the star forming disk into the halo and, if sufficiently
energetic, into the circumgalactic medium (region iii). (b) Correlated feedback from massive young stars overlaps to drive the formation of hot bubbles of X-ray emitting and ionised gas
that correspond with holes in the cooler atomic hydrogen gas. Starlight is shown in white, infra-red observations in red ($T\sim$ 300-500 K),
ionised hydrogen is shown in orange ($T \sim 10^5$ K) and x-ray observations are shown in blue ($T \sim 10^6$ K). M82 image credit:
X-ray: NASA/CXC/JHU/D.Strickland; Optical: NASA/ESA/STScI/AURA/The Hubble Heritage Team; IR:
NASA/JPL-Caltech/Univ. of AZ/C. Engelbracht. 30 Doradus image credit: X-ray: NASA/CXC/PSU/L.Townsley et al.;
Optical: NASA/STScI; Infrared: NASA/JPL/PSU/L.Townsley et al.}
\label{fig:m82}
\end{figure}

\subsection{From star forming regions to `superbubbles' to galactic winds}\label{sec:outflows}

Galactic winds are galaxy-scale gas-outflows composed primarily of hydrogen, but also containing heavier elements and even molecules (see \S\ref{sec:winds}). They are produced from one of two mechanisms, Active Galactic Nuclei feedback (which is the subject of a companion review) and stellar feedback, which is the focus of this perspective. In this section, we combine observations of star forming regions, the ISM and the circumgalactic medium (CGM) of dwarf galaxies to show how stellar feedback drives the formation of `superbubbles' and galaxy-wide winds (see Fig. \ref{fig:m82}).

Stellar feedback begins on 10-100\,pc scales inside young star clusters \citep[e.g.][]{krumholz19}. There, massive stars deposit energy and momentum into the ISM via photoionising feedback and associated radiation pressure \citep[e.g.][]{lopez11,Lopez14}, stellar winds \citep[e.g.][]{stevens03,smith14}, cosmic rays \citep[e.g.][]{Kavanagh20} and SN \citep{smith14}, with additional sources of feedback coming also from binary stars \citep[e.g.][]{smith14}. Observations of nearby star forming regions demonstrate that radiative feedback and winds act to `pre-process' the surrounding birth gas cloud, ionising the surrounding gas to create expanding HII regions \citep[e.g.][]{lopez11,Lopez14,stevens03}. After a few million years, the most massive O, A and B stars explode as Type II SN \citep[e.g.][]{Hopkins12,Agertz13,smith14}. The correlated SN energy from multiple young stars impacts the surrounding, pre-processed, ISM to drive a hot, expanding, bubble of X-ray emitting and ionised gas (\cite{Keller14,Kavanagh20}; and see fig.~\ref{fig:m82}b). As multiple bubbles break out from their natal birth clouds, they expand to sizes >100\,pc and overlap to form `superbubbles'. Several such superbubbles have been observed in exquisite detail in the nearby Large Magellanic Cloud (e.g. \cite{oey02}; and see Fig. \ref{fig:m82}a~i). The largest of these require 100--1000's of supernovae explosions to power them \citep[e.g.][]{martin98,dehorta14}, though other forms of feedback like stellar winds and radiation pressure contribute to the energy budget too, especially at early times before supernovae go off \citep[e.g.][]{Lopez14}. (Note that the terms `bubble' and `superbubble' are sometimes used interchangeably in the literature, while `shell' can refer to the boundary of a bubble, a superbubble or even a larger-scale galactic wind. To avoid confusion, we will define here `bubble' to mean an expanding shell of interstellar material surrounding hot X-ray emitting gas that is $<100$\,pc in diameter. A `superbubble' is then a bubble that is >100\,pc diameter \citep{Kavanagh20}. We will call the shell of the (super)bubble the `(super)bubble shell', while we will refer to the boundary between a galaxy-wide wind and the ISM/CGM as the `wind shell'.)

\begin{figure}[ht]
\centering
\includegraphics[width=\linewidth]{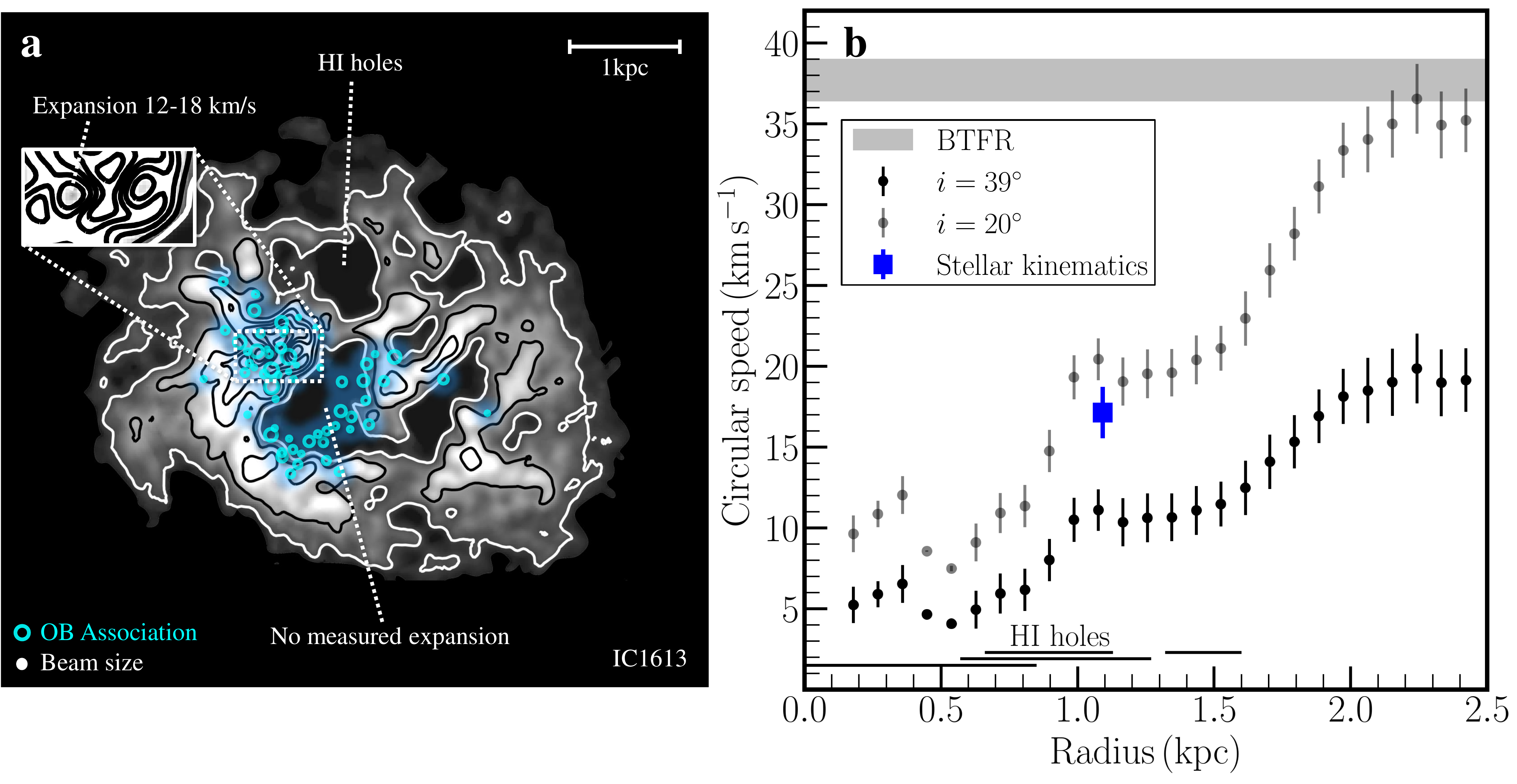}
\vspace{-5mm}
\caption{{\bf Neutral hydrogen (HI) image of the nearby dwarf irregular galaxy, IC1613, showing its prominent HI holes (a) and how these holes impact its derived HI rotation curve (b).} {\bf Panel a:} Isodensity contour map of the HI gas distribution (grey contours), showing both large HI holes that are not expanding and smaller holes (inset panel) that are rapidly expanding. The open blue circles show star forming regions (OB associations). Notice that the locations of the OB associations correlate well with the smaller, rapidly expanding, holes, consistent with the holes owing to stellar feedback. Data taken from \cite{silich06,borissova04}. {\bf Panel b:} Rotation curve for IC1613 derived from its HI gas kinematics (black circles, assuming an inclination of 39$^\circ$), and grey circles (assuming an inclination of 20$^\circ$); \cite{read16b}). The blue square shows the circular speed at the half-light radius, derived from IC1613's stellar kinematics \cite{kirby14}. In all cases, the error bars mark the 68\% confidence intervals. The horizontal grey band marks the expected peak circular speed if IC1613 were to lie along the `Baryonic Tully-Fisher Relation' (BTFR) -- a tight empirical relation between the peak rotation speed of dwarfs and their total baryonic mass \citep[e.g.][]{lelli16,ponomareva21}. The location of the largest HI holes (see panel a) is marked by the black horizontal lines. Notice that the HI holes coincide with prominent `dips' in the rotation curve, while the expansion velocity of the smallest HI holes (inset in panel a) is comparable to the rotation speed in the inner regions. The presence of these holes makes it especially challenging to inclination correct this galaxy and derive its true circular speed curve \citep{read16b}. Standard techniques favour an inclination of $i = 39 \pm 2^\circ$ and a very low inner rotation speed (black data points). This is, however, inconsistent with the circular speed derived from IC1613's stellar kinematics (blue data point). This suggests that IC1613's true inclination is closer to $i \sim 20^\circ$ (grey data points), causing it to no longer be an outlier from the BTFR (grey band).}
\label{fig:holes}
\end{figure}

\begin{figure}[ht]
\centering
\includegraphics[width=\linewidth]{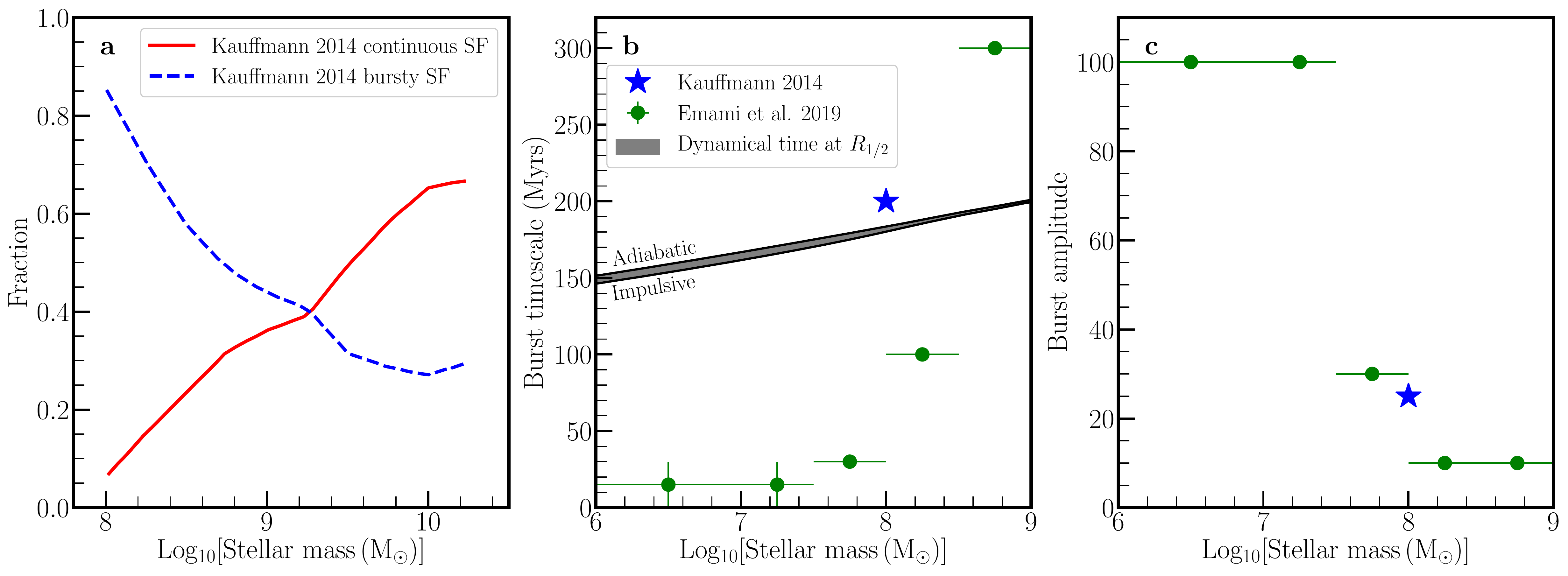}
\caption{{\bf Latest observational constraints on bursty star formation in dwarf galaxies.} {\bf Panel a:} The Star Formation Rate (SFR) density fraction of dwarf galaxies in the Sloan Digital Sky Survey (SDSS) with continuous (red) versus bursty (blue dashed) star formation. {\bf Panel b:} Constraints on the burst timescale ($\tau$ in equation \ref{eqn:burstmodel}) as a function of stellar mass. The green circles show data from \cite{emami19}, where the horizontal error bars mark the bin size and the vertical error bars mark the formal 68\% uncertainties. The blue star shows a back-of-the-envelope estimate of the burst timescale for dwarfs with $M_* \sim 10^8$\,M$_\odot$ from \cite{kauffmann14}. The grey band shows the expected range of dynamical times at the half light radius, $R_{1/2}$, for dwarf irregular galaxies in a $\Lambda$CDM cosmology (see text for details of how this is calculated). Bursts below this band will drive a damaging, impulsive, response on their host galaxy stars and dark matter; bursts above this line yield a more gentle adiabatic response. {\bf Panel c:} Constraints on the maximum amplitude of star formation bursts ($A$ in equation \ref{eqn:burstmodel}) as a function of stellar mass. The green circles and blue star are as in panel b.}
\label{fig:bursty}
\end{figure}

The inverse of superbubbles are `HI holes' -- evacuated regions in the colder HI gas of dwarf galaxies inside which superbubbles typically expand \citep[e.g.][]{oey02,pokhrel20}. The origin of such holes has been debated historically, with theories ranging from gravitational instability \citep{dib05} to gamma-ray bursts and the impact of high velocity clouds \citep[e.g.][]{rhode99}. However, modern data have uncovered a tight correlation between the smaller, faster expanding, HI holes and OB associations -- the sites of massive star formation (e.g. \cite{silich06}, \cite{borissova04}, \cite{pokhrel20} and see Fig. \ref{fig:holes}). This, combined with the fact that the energy required to produce the observed expansion velocity of HI holes -- typically $\sim 10-20$\,km/s -- is consistent with the expected supernova energy input from these OB associations \citep[e.g.][]{pokhrel20}, suggests that stellar feedback is primarily responsible for HI holes in dwarf galaxies. We show an example of HI holes and their correlation with OB associations for the nearby dwarf irregular galaxy IC1613 in Fig. \ref{fig:holes} (left panel; data taken from \cite{silich06,borissova04}). Notice that the smallest, fastest-expanding, holes correlate well with OB associations (blue), while the largest HI holes have no measured expansion and correlate poorly with the OB associations. This latter provides further evidence that the HI holes originate from expanding hot superbubbles. Assuming an initial expansion velocity of $v_i = 20$\,km/s (typical for the smallest holes; \cite{pokhrel20}), an initial hole diameter of $r_i = 100$\,pc, a final diameter of $r_f = 1$\,kpc, and that the hole expansion velocity linearly falls to zero as the hole reaches its maximum size, an HI hole will reach its maximum diameter in $t_f = 2(r_f-r_i)/v_i \sim 88\,{\rm Myrs}$. For a typical dwarf irregular galaxy, this is $\sim$half of the local dynamical time (Fig. \ref{fig:bursty}, middle panel). As such, we expect that any correlation between local star formation and HI holes will be rapidly erased as the hole expands, the young stars die and any remaining stars begin to phase mix in the disc.

As superbubbles continue to grow, they can overlap to form galaxy-wide winds (fig.~\ref{fig:m82}a~ii). These winds remove gas from the star forming disk and, if sufficiently energetic, into the CGM (fig.~\ref{fig:m82}a~iii and see \cite{zheng19}). By measuring the dynamical ages and expansion times of winds and comparing to the recent star formation activity of the galaxy, we can show the winds are consistent with being launched by starburst activity. For example, using H$\alpha$ imaging and spectroscopy \cite{westmoquette08} measured a dynamical age of $\lesssim25$~Myr for the galactic outflow in NGC 1569, consistent with recent star formation activity. A number of other studies find similar timescales \cite[e.g.][]{martin98,mcquinn18}.

\subsection{Dwarf galaxy winds}\label{sec:winds}

The presence of large-scale gaseous outflows have long been inferred from observations of dwarf galaxies and an exceptional example is the majestic M82 system (\cite{lynds63}; and see Fig.~\ref{fig:m82}), the only dwarf galaxy in the local Universe that exhibits a developed and powerful wind. While something of an outlier, M82 helps to demonstrate the possible characteristics of an extreme wind. Galactic winds are cone-like in shape, moving at velocities 25-100~kms$^{-1}$ perpendicular to the dwarf galaxy's stellar and gaseous disc \cite{martin98,schwartz04, mcquinn19}. They are also multi-phase, typically containing hot ($T>10^5\,{\rm K}$), warm ($10^3<T<10^5\,{\rm K}$) and cold ($T<10^3\,{\rm K}$) components that entrain metals, molecules and dust \cite[e.g.][]{veilleux05,martin06}. While M82 presents us with a very obvious detection of winds, identifying and catagorising winds in dwarf galaxies is observationally challenging. The winds are diffuse, and the irregular morphologies and kinematics make characterisation of winds difficult. For the nearest systems, clear H$\alpha$ shells and bubbles can be identifed e.g. \cite{martin98}, but further afield, classifying and characterising these winds is more subjective (see e.g. \cite{mcquinn19} for a discussion).
Hard X-ray observations constrain the hottest phase of the wind ($T > 10^6$\,K), and are usually concentrated on the starburst generating the wind as this is where the denser gas is found (\cite{martin02,strickland09}; Fig.~\ref{fig:m82}b). Soft X-ray observations probe gas locked in the slightly cooler $\sim10^6$\,K phase. This is usually observed at the boundary of the hot and warm phases, and traces (super)bubbles formed from correlated SNe events (\S\ref{sec:outflows} and Fig.~\ref{fig:m82}a~iii). The hot phase is thought to be the most energetic and most metal-enhanced component. It has the greatest chance of escaping the disk and even halo of the galaxy, enriching the CGM \citep{zheng19}, as shown for M82 \cite{strickland09}. The presence of diffuse soft X-ray emission beyond the HI disk is thus typically interpreted as an outflow event. Unfortunately, due to the low surface brightness of the emission, X-ray observations (both hard and soft) are typically limited to the most energetic starburst systems (such as M82, NGC~4449, NCG~625, NGC~1569, NGC~4214, \cite{martin98,martin02,summers03,mcquinn18}), and are likely to fade just $\sim25$\,Myr after the starburst that generated them \cite{mcquinn18}. As such, we do not expect such a hot wind to be ubiquitous in dwarfs. 

The warm phase ($T\sim10^5\,{\rm K}$) is traced with ionised hydrogen using observations in the far UV \cite[e.g.][]{heckman15,chisholm17} and optical/H$\alpha$ \cite[e.g.][]{meurer92,martin98,mcquinn19}. Most of the mass is entrained in this phase (fig.~\ref{fig:m82}a~i,ii), and so it is crucial for determining the rate of mass outflow -- $\dot{M}_{\rm o}$ -- and the mass-loading of the winds (\S\ref{sec:massload}). Rest-frame UV absorption lines are commonly used to study outflows in galaxies at high and low redshift using space-based observatories such as HST-COS and the former FUSE spacecraft \cite[e.g.][]{veilleux05,heckman15,chisholm17,chisholm17b}. In general, UV and H$\alpha$ studies find that dwarf galaxies with more centrally concentrated star formation are more likely to host a significant wind \cite[e.g.][]{mcquinn19}. They also find that the majority of material in dwarf galaxy winds are not moving fast enough to escape the gravitational potential. Instead, the winds are consistent with moving the majority material into the halos of the dwarf where it can be recycled, while a smaller fraction can be transported into the intergalactic medium (IGM).

Finally, the cold phase ($T\sim10^4\,{\rm K}$) is typically traced with HI, H$_2$  and CO observations observations (fig.~\ref{fig:m82}b). The molecular phase is thought to carry a significant mass fraction in outflows \cite[e.g.][]{krieger19}. Cold outflows have been observed in some dwarf systems \cite[e.g.][]{lelli14a}, including M82 \cite[e.g.][]{leroy15,krieger21} where CO and $H_2$ are associated with the hot-outflow, and can be traced to distances of $\sim3\,$kpc beyond the mid-plane. But the neutral component of winds is expected and typically observed to be small. Nonetheless, it is important as a key discriminator of theoretical models, many of which predict purely hot outflows for dwarfs  \citep[e.g.][]{pandya21}. It may point, for example, to an important role for cosmic ray feedback. Cosmic rays accelerate gas more gradually and at all locations around the disc, not just near star forming regions, yielding cooler winds \citep[e.g.][]{Booth13}.

\subsubsection{The link between star formation and outflow rates}
\label{sec:energetics}

Winds and outflows are ultimately powered by star formation activity (\S\ref{sec:outflows}). Knowing the SFR during a starburst is key for characterising an observational wind and quantifying the mass loss so it can be compared to theoretical predictions. This is encapsulated in the mass loading factor, $\eta$, which we discuss in \S\ref{sec:massload}. This compares the outflow rate, $\dot{M}_{\rm o}$, to the instantaneous star formation rate (SFR). As such, a solid understanding of the SFR is essential for understanding the outflow rates of galactic winds. 

There are several techniques for determining the SFR. For nearby galaxies where we can resolve individual stars, their stellar populations are used to  determine the star formation history (SFH) over periods of hundreds of Myr to a Hubble time using the synthetic colour magnitude diagram (CMD) fitting approach \cite[e.g.][]{tolstoy09,weisz11,weisz12,weisz15}. In this process, a Monte Carlo calculation is run to reproduce a CMD's features using stellar models (isochrones) with different ages, initial mass functions (IMF), binary fractions, metallicity and other physics to find out which combination of stellar populations best match the observations \cite[e.g.][]{weisz11,brown14,weisz14a,weisz14b,deboer14,mcquinn10a, gallart21,rusakov21}. For starbursting dwarfs with resolved bright stars, one can well-constrain the SFH over the past $\sim1-2$~Gyr, covering their recent and ongoing starbursts \cite[e.g.][]{mcquinn10a,mcquinn10,mcquinn19}. For systems where the stars cannot be resolved, one can infer recent SFRs using H$\alpha$ or UV observations. As H$\alpha$ traces O stars, the strength of the flux equilibrates on timescales of $\sim5$~Myr during constant star formation. UV observations trace O, A and B stars. As A and B stars are longer lived, the time resolution for these observations is $\sim100$~Myr. As SFRs can vary by a factor of a few on timescales shorter than 100~Myr, UV observations can be uncertain by a similar factor \cite{mcquinn10a}.

With a measure of the SFR during a wind-generating starburst, one can then better characterize the mass outflow, and compare with theoretical expectations. We discuss this in detail in the following subsection.

\subsubsection{The wind mass-loading factor}
\label{sec:massload}

\begin{figure}[ht!]
\centering
\includegraphics[width=\linewidth]{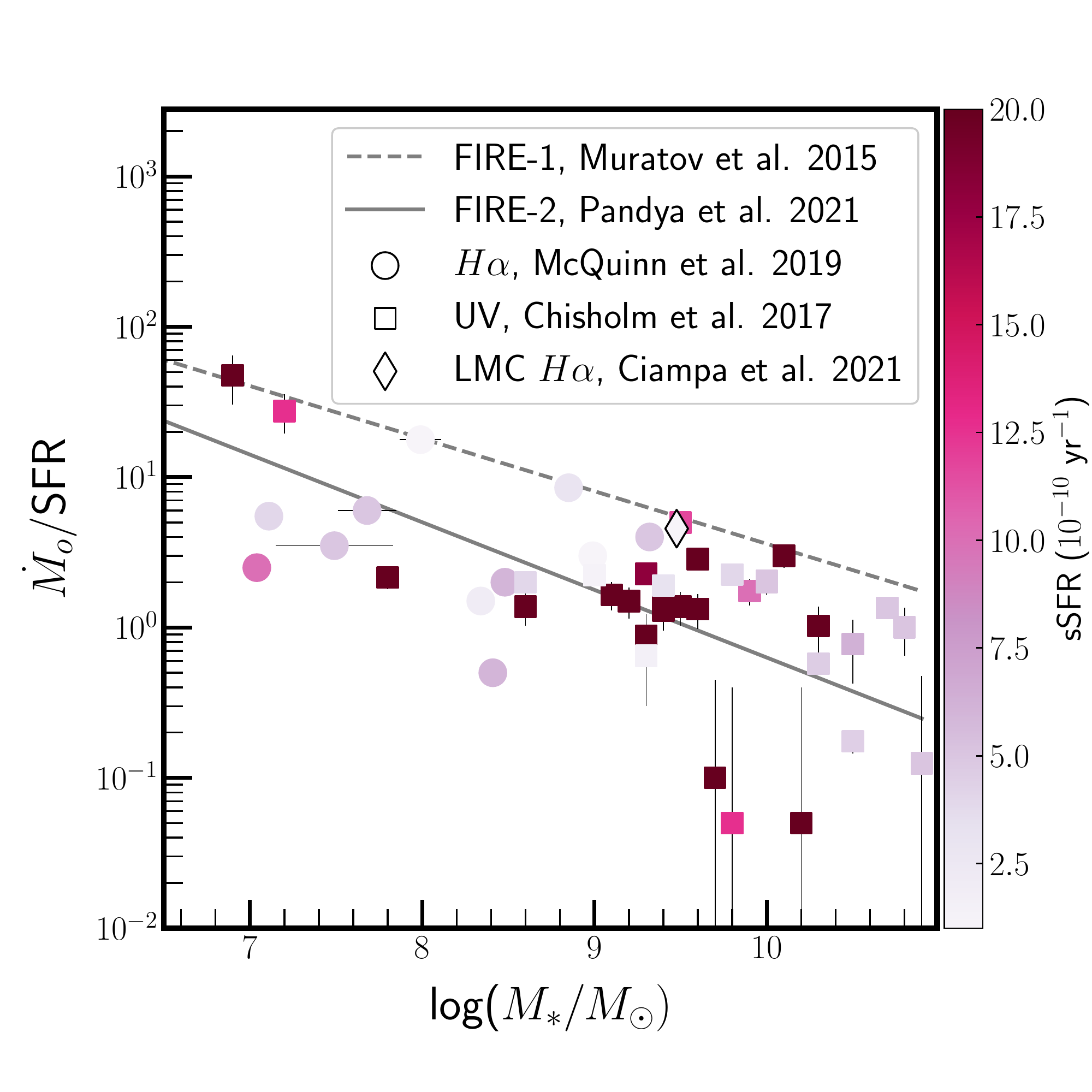}
\caption{{\bf Mass loading rates ($\dot{M}_{\rm o}/{\rm SFR}$) as a function of stellar mass from several recent studies of outflows in dwarf galaxies, colour-coded by their specific star formation rate}. All points are plotted with uncertainties, ubt some are smaller than the size of the markers. The dashed line shows a theoretical prediction for mass loading taken from the FIRE-1 simulations, which is typically higher than the majority of observations. The solid line shows the same but from the FIRE-2 simulations, which still over-predicts the mass-loading at the lowest masses. Both are similar to other models (a more full comparison can be seen in \cite{mcquinn19}). Intriguingly, the recent measure of mass-loading from the LMC from deep H$\alpha$ mapping is consistent with these models.}
\label{fig:massloading}
\end{figure}

In \S\ref{sec:winds}, we discussed how H$\alpha$ and far-UV can be used to measure the mass entrained in galactic winds ($\dot{M}_{\rm o}$), while in \S\ref{sec:energetics} we gave an overview of the star formation metrics used to determine the available power provided by the stars. Here, we combine these to determine the `strength' of a galactic winds using the mass-loading factor, $\eta$, defined as the ratio between the rate of mass ejection in the winds $\dot{M}_{\rm o}$ and the star formation rate which powers the wind ($\dot{M}_*$, or SFR).

Several recent works have reported mass loading values for dwarf galaxies spanning $10^7\lesssim M_*\lesssim 10^{10}\,{\rm M}_\odot$, \cite[e.g.][]{heckman15,chisholm17,mcquinn19,ciampa21}. These are shown as a function of stellar mass in fig.~\ref{fig:massloading}. These studies also investigate whether $\eta$ scales with stellar mass, circular velocity (a proxy for total mass), SFR and spatial concentration of star formation. There are differences in their findings, but some of these are due to subtleties in observational methods, galaxy samples and modelling assumptions. They also compare their findings to high resolution simulations with different `subgrid' models for star formation and stellar feedback. We summarise these findings here, and discuss their implications.

Given the shallower potential wells of dwarf galaxies, it is typically assumed that $\eta$ will increase with decreasing mass or luminosity. But this is not always seen in observations. We show three compilations of data in fig.~\ref{fig:massloading}: UV samples from \cite{heckman15} and \cite{chisholm17}, and H$\alpha$ results from \cite{mcquinn19}. Interestingly, \cite{heckman15}and \cite{mcquinn19} find little to no relation between stellar mass (or total mass) and $\eta$, while \cite{chisholm17} find a reasonably steep relation for their sample of 7 systems.

While these results seem contradictory, there are some nuances. For the two UV samples, different assumptions are used for the outflow density, metallicity and spatial coverage. \cite{heckman15} assume a constant density and metallicity across their sample, and set the outflow radius to twice that of the star formation region, while \cite{chisholm17} uses observationally motivated values on a galaxy by galaxy basis. They find that this leads to values for $\eta$ that differ by a factor of 10, c.f. \cite{heckman15}. In \cite{mcquinn19}, they use deep H$\alpha$ maps for their systems, meaning they are able to directly measure the extent of the winds, an advantage over pencil beam UV surveys which have a limited coverage of the wind region.

The way that star formation rate is measured also differs between studies. The H$\alpha$ study from \cite{mcquinn19} use the temporally resolved star formation history derived from HST photometry, whereas the UV studies scale the UV flux. This has the effect of tying the SFR to a 100 Myr timescale (longer than the warm gas cooling time). These also depend on scaling relations which can under-predict the SFR by 50\%, compared to those measured empirically from colour magnitude diagrams, leading to values for mass loading that are 50\% higher \cite{mcquinn15b}.

Additionally, the properties of the sample galaxies differ. In fig.~\ref{fig:massloading}, we colour-code each point by its specific star formation rate (sSFR) to show how extreme the current star formation is in each system. In general, the UV studies include more extreme starbursts than the H$\alpha$ study, which could lead to different biases for both methods. Plus, all three studies find that it is not only the {\it amount} of star formation in these systems that matters, but its spatial distribution. Systems with strongly concentrated starbursts show higher mass loss rates than those with more extended star formation. 

Comparing relations with total mass (or circular velocity) between observations and simulations can be misleading, as one is not comparing like with like. The circular velocities can be measured at vastly different radii, and the maximum circular velocity is not always observed as the rotation curves are still rising within the extent of the dataset. There are also issues with non-circular motions in HI gas and measuring precise inclinations for galaxies which can introduce biases (see \S\ref{sec:morph}). Comparing to stellar mass is thus more straightforward. In fig.~\ref{fig:massloading} we show the mass-loading prediction within the FIRE-1 and FIRE-2 suite of simulations \cite{muratov15,pandya21} which increases with decreasing stellar mass. The two versions span the observations, with FIRE-1 typically higher than the majority of observations at all masses, but the difference is particularly striking below $M_*<10^9\,{\rm M}_\odot$. FIRE-2 is closer to the bulk of observations, although it still over-predicts at the lowest masses. This is seen in other large-scale simulations (as seen in fig.~10 of \cite{mcquinn19}). So, are cosmological simulations overestimating the amount of mass loss in the lowest mass galaxies?

Maybe, maybe not. Higher resolution simulations of individual dwarf galaxies do predict lower mass-loss rates than cosmological simulations discussed above, more comparable to observations \cite[e.g.][]{hu17,hu19,smith21}. These show the importance of different modes of stellar feedback, and their impact on mass-loading. A further consideration is whether observations are capturing {\rm all} of the material within winds in low mass systems. As all authors note, these features are extremely challenging to observe. If there is more material just below the detection threshold of these observations, the mass-loading factors could rise significantly. And this has been seen in a new study of the LMC using extremely deep H$\alpha$ observations across the whole galaxy \cite{ciampa21}. The LMC has a lower sSFR than the star-bursting dwarfs presented in most other studies, and so may be assumed to have a lower mass-loading factor. However, in fig.~\ref{fig:massloading} (where the LMC is shown as a diamond) it is perfectly consistent with the higher FIRE-1 predictions. While this result is for one galaxy, which also has a range of assumptions made about the geometry and distribution of the wind, it makes a compelling case that we may not be capturing all the outflowing material in more distant galaxies. Future facilities will allow us to obtain deeper data for more systems to test this.

\section{The impact of feedback on dwarf galaxies}\label{sec:impact}

\subsection{Bursty star formation}\label{sec:sfh}

Stellar feedback acts to regulate star formation in dwarfs, lowering their stellar masses significantly as compared to a Universe in which no feedback were present \citep[e.g.][]{dekel86,read19b}. It also impacts the whole baryon cycle and, therefore, how star formation proceeds with time. The latest numerical simulations of dwarf galaxy formation have now reached a resolution and physical fidelity that can resolve the cold, dense, ISM, with a mass and spatial resolution of better than $100-1000$\,M$_\odot$ and $\Delta x \sim 10-100$\,pc, gas cooling below $T \sim 100$\,K, and resolved gas densities of $\rho_g > 100$\,atoms/cc \citep[e.g.][]{mashchenko08,teyssier13,wise14,onorbe15,read16,agertz20,jahn21,gutcke21a}. This is a key milestone since it means that simulations can capture the formation of the most massive molecular clouds, forming stars in the right places at the right times and, thereby, having feedback impact the ISM in a more realistic manner than the previous generation of simulations (see the companion perspective article on simulations of dwarfs). Such simulations universally predict that star formation in dwarfs should be `bursty', with a duty cycle comparable to the local dynamical time and peak-to-trough variations in the star formation rate of a factor $\sim 5-10$ \citep{teyssier13}. As such, the variability of star formation in dwarfs serves as an indirect probe of feedback and an important test of the latest theoretical models.

In Figure \ref{fig:bursty}, we collate the latest data constraints on the variability of star formation in dwarf galaxies. Such variability is likely driven by a range of physical processes, including stellar feedback, galactic interactions (e.g. \cite{noel09,deboer14}) and galaxy mergers \citep[e.g.][]{adamo11,kimbro21}. However, galactic interactions and mergers typically occur on $>$Gyr timescales, much longer than the dwarf starburst cycle \citep[e.g.][]{teyssier13}. As such, while variability on $\sim$Gyr timescales has been reported for nearby dwarfs  \citep[e.g.][]{mcquinn10a,mcquinn10}, we focus here on studies that probe the shortest period bursts as these are most likely to probe the starburst cycle. The age-resolution of star formation histories derived for individual galaxies from resolved colour magnitude diagrams is $\sim$\,1Gyr, falling to $\sim 200$\,Myrs only for the youngest-age stars (see \S\ref{sec:energetics} and e.g. \cite{noel09,mcquinn10a,mcquinn10,weisz11,brown14,weisz14a,gallart21,rusakov21}). Since this is coarser than the burst timescale predicted by the latest simulations, we must take a different approach: studying the statistics of star formation indicators that probe different timescales (see \S\ref{sec:energetics}) across whole dwarf galaxy populations \citep[e.g.][]{glazebrook99}. Using data for 482,755 dwarf galaxies selected from the Sloan Digital Sky Survey (SDSS), \cite{kauffmann14} took exactly this approach. They probed $>$100\,Myr timescales using the 4000 angstrom break in the galaxy spectrum (that is only present if there are no/few hot young stars) and shorter timescales using H$\alpha$ lines (that probe ionising radiation from hot young stars). They fitted the data with a continuous star formation model, one with a recent burst in the past 100\,Myrs, and one with continuous bursts. More recently, \cite{emami19} performed a similar study using 1300-2000 angstrom far-UV continuum and H$\alpha$ photometric data drawn from the 11HUGS survey \citep{kennicutt08,lee09}. They fitted an exponential burst model to the data:

\begin{equation} 
{\rm SFR}(t) = \left\{
\begin{array}{ll}
e^{t/\tau}, & {\rm if} \,\,\,\, 0 \le t < D \\
e^{-(t-2D)/\tau}, & {\rm if} \,\,\,\, D < t \le 2D
\end{array}
\right.
\label{eqn:burstmodel}
\end{equation} 
where $\tau$ is the `burst timescale', $D$ is the `burst duration' and the burst is assumed to repeat periodically. The `burst amplitude', $A$, then follows as: $A = e^{D/\tau}$.

The results from these two studies are reproduced in Fig. \ref{fig:bursty}. The left panel shows the Star Formation Rate (SFR) density fraction of dwarfs with continuous star formation (blue) versus those currently undergoing a burst (red) drawn from the \cite{kauffmann14} study. Notice that the burst fraction approaches unity as the stellar mass is decreased (recall that we define dwarf galaxies to have $M_* < 10^9$\,M$_\odot$). The middle panel shows the `burst timescale' (defined as in equation \ref{eqn:burstmodel}).
\cite{kauffmann14} only provide a back-of-the-envelope estimate of the burst timescale for dwarfs with $M_* \sim 10^8$\,M$_\odot$ (blue star, middle panel of Figure \ref{fig:bursty}). Nonetheless, this is in excellent agreement with the more recent analysis by \cite{emami19} (green data points). The grey band marks the expected range of dynamical times at the half light radius, $R_{1/2}$, for dwarf irregular galaxies in a $\Lambda$CDM cosmology. (For this calculation, we use the $M_*-M_{200}$ relation from \cite{read17}, and its $1-\sigma$ scatter, to estimate the dark matter halo mass as a function of stellar mass. We then assume the median halo concentration parameter from the \cite{dutton14} $M_{200}-c_{200}$ relation, valid for a $\Lambda$CDM cosmology. We estimate the stellar half mass radius, $R_{1/2}$, using the $R_{1/2}-r_{200}$ relation from \cite{kravtsov13}. And, finally, we assume that the dynamical mass is dominated by the dark matter halo which is of \cite{navarro96} form.) If the burst timescale is longer than the local dynamical time, then the stars and dark matter in the galaxy will respond adiabatically to any gas mass loss driven by the burst. By contrast, if the burst timescale is shorter than the local dynamical time, this will yield an impulsive response. This is an important distinction because an impulsive response will irreversibly pump energy into the orbits of stars and dark matter particles, lowering the inner dark matter density \citep[e.g.][]{navarro96b,Read05DMheats,pontzen12}. We will discuss this further in \S\ref{sec:dm}. 

Finally, the right panel of Fig. \ref{fig:bursty} shows observational constraints on the burst amplitude, $A$ (defined as in equation \ref{eqn:burstmodel}), from \cite{emami19} (green data points). As in the middle panel, an estimate from \cite{kauffmann14} is marked by a blue star. Notice the good agreement between \cite{kauffmann14} and \cite{emami19} despite the fact that these studies use very different data and methodologies. Notice further that the burst amplitude increases dramatically as $M_*$ falls.

The above data demonstrate that star formation in dwarf galaxies is indeed bursty. However, it is important to note that the precise burst frequency, amplitude and duration may be impacted by systematic errors. Firstly, near the flux-limit, surveys will be biased towards star-bursting dwarfs because they are more luminous \citep[e.g.][]{dominguez15}. Secondly, the combination of UV continuum and hydrogen recombination lines is an imperfect measure of starbursts \citep[e.g.][]{dominguez15,sparre17,emami19}. Thirdly, simple `exponential' burst models, or similar, may produce biased inferences when applied to real star formation histories \citep[e.g.][]{emami19}. Mock data drawn from the `FIRE-2' simulations suggests that only the first of these three is a significant concern \cite{emami19}. Detailed comparisons between these same simulations and the data from \cite{emami19} (Figure \ref{fig:bursty}) shows good agreement in the burst properties of low mass dwarfs, but above $M_* > 10^8$\,M$_\odot$, the simulated burst timescale is too short ($<30$\,Myrs). The reasons for this discrepancy remain to be understood, but could plausibly be explained by missing `sub-grid' physics and/or resolution effects \citep[e.g.][]{applebaum20,smith21,prgomet21}.

\subsection{Quenching}\label{sec:quenching}

While star forming dwarf irregular galaxies show a continuous cycle of star formation bursts, dwarf spheroidal galaxies show no recent star formation and no detectable HI gas \citep[e.g.][]{mcconnachie12,Simon19}. This `quenching' could be another hallmark of the impact of stellar feedback. However, dwarf spheroidals are typically found close to a larger host galaxy while star forming irregulars are more isolated \citep[e.g.][]{blitz00,grebel03}. Indeed, \cite{geha12} find that essentially no dwarf galaxies -- by our definition of a dwarf ($M_* < 10^9$\,$M_\odot$) -- are quenched if sufficiently isolated. This suggests that most dwarfs are primarily quenched by environmental effects, though these are likely enhanced by stellar feedback \citep[e.g.][]{gatto13}. For more massive dwarfs, the most likely environmental mechanism is `ram-pressure' stripping of their interstellar medium when they fall into a larger host galaxy \citep[e.g.][]{gatto13}; for the very smallest `ultra-faint' dwarfs ($M_* \lesssim 10^{5-6}$\,M$_\odot$), the most likely mechanism is reionisation -- ionising photons from external galaxies and quasars \citep{efstathiou92} in the redshift range $z \sim 6-10$ \citep[e.g.][]{bouwens15}. Evidence for this latter comes primarily from ultra-faint dwarf galaxy star formation histories, as measured from deep colour magnitude diagrams. \cite{brown14} study six nearby ultra-faints, finding that all have extremely similar stellar populations,  with close to 100\% of their stars formed by $z=4$. This is consistent with the latest models of reioinisation quenching in which self-shielding of dense gas allows some star formation to proceed even after reionisation has completed \citep[e.g.][]{onorbe15,jeon17,agertz20,katz20,rey20}. Further evidence for this comes from the dynamical masses of ultra-faint dwarfs. Estimates from both their stellar kinematics and abundance matching place ultra-faints in halos of mass $M_{200} \sim 10^{9}$\,M$_\odot$ \citep[e.g.][]{read19b}, consistent also with the latest models of reionisation quenching \citep[e.g.][]{onorbe15,jeon17,agertz20,mina20,katz20,gutcke21b}.

 Despite the important role of environment quenching, it is possible that some dwarf spheroidals are quenched by their own internal stellar feedback alone \citep[e.g.][]{dekel86,gallart21}. Such dwarfs should have more extended star formation than the ultra-faints and yet be as isolated as the star forming dwarf irregulars. An unambiguous candidate has yet to be found \citep[e.g.][]{mcconnachie12,gallart21}, but may be discovered in upcoming surveys like the Vera Rubin Observatory.

The suppression and quenching of star formation in dwarf galaxies is key to solving a long-standing tension between the number of observed galaxies orbiting the Milky Way and M31 and the large numbers of bound dark matter halos predicted in $\Lambda$CDM structure formation simulations -- the `Missing Satellites Problem' (MSP; e.g. \citep{moore99,klypin99,bullock17}). The scale of this problem has reduced with time, as newer predictions have revised down the expected number of satellites and new surveys have found ever larger numbers of nearby ultra-faint dwarfs \citep[e.g.][]{simon07,Simon19}. The latest observations of isolated dwarf irregulars demonstrate that star formation becomes increasingly inefficient as the halo mass is reduced, almost certainly due to stellar feedback \citep[e.g.][]{Read05,read17,katz17,posti19}. This solves the MSP, at least down to the mass-scale of ultra-faint dwarfs, by rendering the lowest mass dark matter halos too faint to be observed, or even fully dark \citep[e.g.][]{simon07,read17,aubert18,jethwa18,kim18,read19b,Nadler20}. (Note that isolated dwarf irregulars are a better test-bed for studying feedback processes than satellite dwarf galaxies. This is because satellites can lose significant gas, stellar and dark matter mass and/or have their star formation enhanced or shut down via environmental processes \citep[e.g.][]{read17,read19b}. This makes it much harder to determine which aspects of their observed properties today owe to stellar feedback alone.)

\subsection{Morphology \& size}\label{sec:morph}

If bursty star formation drives repeated outflows on a timescale comparable to, or shorter, than the local dynamical time then this will drive an irreversible, impulsive, energy injection into the stars, dark matter and gas in the host dwarf \citep[e.g.][]{navarro96b,Read05DMheats,pontzen12}. As shown in Fig. \ref{fig:bursty} (middle panel), current observational constraints on star formation in dwarfs suggests that this is indeed the case. If so, then stellar feedback should also impact the size and morphology of dwarf galaxies. There are several lines of observational evidence for this, albeit each one indirect. Firstly, nearby dwarf irregulars are observed to show a radial gradient in their stellar ages, with older stars lying further out and younger stars being more centrally concentrated \citep{zhang12}. \cite{elbadry16} show that this is expected if star formation proceeds in repeated impulsive bursts, driving gas outflows that cause stars to slowly migrate outwards. Secondly, the same process drives size-fluctuations that are consistent with current observational constraints \citep{emami21}. Thirdly, it can explain the existence of `ultra-diffuse' low surface brightness dwarfs (defined as galaxies with large effective radii, over 1.5~kpc, and low central surface brightnesses, less than $\mu_{0,V}\sim23.5$, \cite{vandokkum15}) \cite{dicintio17,chan18}, though we note that this formation pathway is not without its problems \citep[e.g.][]{chan18,jiang19,kadofong21} and is just one amongst several proposed mechanisms \citep[e.g.][]{liao19,tremmel20}. And, fourthly it yields kinematically hotter stellar populations \citep[e.g.][]{teyssier13}, similar to those observed in nearby dwarf irregulars \citep{leaman12,wheeler17}. 

Stellar feedback can also cause fluctuations in the gas rotation velocity that hamper our ability to accurately mass-model dwarfs. Here, there are two main effects. Firstly, numerical models predict that the gas rotation speed should physically fluctuate, rising more steeply during a star burst and less steeply during quiescent phases \citep{read16b}. Secondly, the presence of expanding HI holes -- that move at a speed comparable to the local rotation curve (see Fig. \ref{fig:holes}; left panel) -- can cause `dips' in the inner rotation curve and large systematic errors in the derived inclination for dwarfs that are close to face-on (with inclination, $i \lesssim 40^\circ$; \cite{read16b}). In Fig. \ref{fig:holes} (right panel), we show an example of this behaviour for the isolated dwarf irregular IC1613. This is a rare case where we have both gas kinematics from the LittleTHINGS survey \citep{oh11} and stellar kinematics \citep{kirby14}. The black data points show the rotation curve for IC1613 derived from its HI gas kinematics using the state-of-the-art tilted ring fitting code $^{\rm 3D}$barolo \citep{iorio17}. $^{\rm 3D}$barolo favours an inclination for IC1613 of $i=39 \pm 2^\circ$ with small formal uncertainties. This yields a very slowly rising rotation curve that, at face value, implies the presence of an enormous low-density dark matter core with little-to-no inner dark matter at all \citep{oman16}. It also makes IC1613 a significant outlier from the `Baryonic Tully Fisher Relation' (BTFR) -- a tight empirical relation between the peak rotation speed of dwarfs and their total baryonic mass \citep[e.g.][]{lelli16,ponomareva21}; the expected rotation velocity for IC1613 if it were to lie on the BTFR is marked by the horizontal grey band. However, notice that IC1613's slowly rising rotation curve is inconsistent with its stellar kinematics (blue data point; and see \cite{oman16}). If we demand that the gas and stellar kinematics are consistent, this yields a much lower inclination of $i \sim 20^\circ$ and a more steeply rising inner rotation curve (grey data points). This moves IC1613 back onto the BTFR, making it consistent with other similar nearby dwarf irregulars \citep[e.g.][]{iorio17}. \cite{read16b} use a simulated mock dwarf to understand this behaviour. They show that if a dwarf is close to face-on and its outer HI isodensity contours are distorted -- which is more likely to happen if the galaxy has undergone a recent starburst -- then $^{\rm 3D}$barolo's tilted ring fit biases the inclination high. This bias grows as the inclination shrinks, and can be much larger than the formal uncertainties for $i < 40^\circ$ -- precisely the situation for IC1613.

The above detailed case-study for IC1613 highlights how feedback can complicate the mass modelling of dwarfs. Yet, due to their greater distance, most nearby dwarfs have far poorer data than are available for IC1613. Few have stellar kinematics while most do not even resolve the inner rotation curve rise nor the peak of the rotation curve at large radii \citep[e.g.][]{oman21}. Constructing mass modelling tools that are robust to the complications induced by stellar feedback will be vital if we are to fully exploit upcoming data for large, volume complete, samples of dwarfs that will come from future surveys with the SKA \citep[e.g.][]{ponomareva21}.

\subsection{The inner dark matter density of dwarfs}\label{sec:dm}

In \S\ref{sec:morph}, we argued that repeated impulsive gas blow-out will pump energy into the orbits of stars, causing dwarf galaxies to slowly expand and puff up vertically. The same physics is expected to act also on the dark matter particle orbits, an effect that has become known as `dark matter heating' \citep[e.g.][]{navarro96,Read05DMheats,pontzen12,read16}. This is of particular importance since it can solve two further small-scale tensions in the $\Lambda$CDM model. The first is the `cusp-core problem': pure-dark matter structure formation simulations in $\Lambda$CDM predict high density central dark matter `cusps' in dwarf galaxies, whereas observations have long-favoured shallower, lower density, `cores' \citep[e.g.][]{flores94,moore94,bullock17}. Dark matter heating alleviates this tension by slowly pushing dark matter out from the centres of star forming dwarfs, transforming the cusp to a core \citep[e.g.][]{Read05DMheats,pontzen12}. The second tension is the `Too Big to Fail' problem (TBTF; e.g. \cite{read06,boylan11,bullock17}): the inner densities of massive satellite dwarfs are also lower than expected from pure-dark matter structure formation simulations in $\Lambda$CDM. When coupled with the fact that tides have a stronger effect on lower density dwarfs, TBTF can be recast as the `cusp-core problem' but applied to satellites \citep[e.g.][]{read06,brooks14,read17}. As such, dark matter heating naturally alleviates TBTF too \citep[e.g.][]{brooks14,read16}. (Note that dark matter heating may be further amplified in dwarf galaxies that host an AGN, if feedback from the AGN drives additional, impulsive, gravitational potential fluctuations \citep[e.g.][]{martizzi13}. It remains to be seen if such effects are required in addition to stellar feedback to explain the observed properties of dwarfs.)

There is mounting observational evidence from the kinematics of stars and gas in dwarfs galaxies that their inner dark matter halos do indeed puff up in response to stellar feedback. A key prediction of dark matter heating models is that complete core formation must take many dynamical times (since each starburst cycle takes $\sim$a dynamical time (Fig. \ref{fig:bursty}), while each cycle couples supernova energy very inefficiently to the dark matter particle orbits \citep[e.g.][]{Read05DMheats,pontzen12,penarrubia12}). As such, many cycles are required to sufficiently lower the inner dark matter density to a flat, constant density, dark matter core \citep{read16}. This means that, all other things being equal, dwarf galaxies with little star formation should have steeper central dark matter cusps than those with more star formation. \cite{dicintio14} parameterise this using the ratio of stellar-to-dark matter halo mass, $M_*/M_{200}$, predicting that galaxies should be cuspy for $M_*/M_{200} \lesssim 5\times 10^{-4}$ and maximally cored for $M_*/M_{200} \sim 5\times 10^{-3}$.

\cite{Read19} measure the inner dark matter densities of 16 nearby dwarf galaxies with high quality data and a range of star formation histories to test this prediction. They find an anticorrelation between the inner dark matter density (measured at 150\,pc) and $M_*/M_{200}$, with the approximate transition from cuspy to cored systems occurring at $M_*/M_{200} \sim 5 \times 10^{-4}$, exactly as predicted by \cite{dicintio14}. These results have been corroborated and expanded upon by \cite{Bouche21}. Their data overlap with \cite{Read19} at $M_*/M_{200} \sim 5\times 10^{-3}$, but extend to higher $M_*/M_{200}$, showing the increasing inefficiency of dark matter heating as $M_*/M_{200}$ rises above $\sim 10^{-2}$, again as predicted by \cite{dicintio14}.

The above observational constraints bring us to the remarkable conclusion that stellar feedback not only regulates star formation in dwarf galaxies, but actively reshapes their inner mass distributions. This implies that, whatever dark matter is, it is dynamical -- it can be moved around by a fluctuating gravitational field. This is suggestive of models in which dark matter is a new, weakly interacting, particle. But the nature of this particle remains elusive, and a wide range of models remain consistent with the data for the time-being \citep[e.g.][]{kaplinghat16,olive16,hoferichter17,du18,boyarski19}.

\subsection{The stellar mass-metallicity relation}
\label{sec:mfeh}

Stellar feedback also regulates the chemistry of dwarf galaxies. Numerical models suggest that metals are more efficiently ejected by winds into the CGM \citep[e.g.][]{maclow99,emerick20,emerick20b}, which is supported by the tentative observation of a metal-rich CGM around the isolated dwarf irregular, WLM \citep{zheng19}. Indeed, \cite{agertz20} argue that the stellar-mass metallicity relation ([Fe/H]-$M_V$) is the only observable, global, scaling relation for nearby dwarfs that is sensitive to feedback physics. 

\begin{figure}[ht]
\centering
\includegraphics[width=\linewidth]{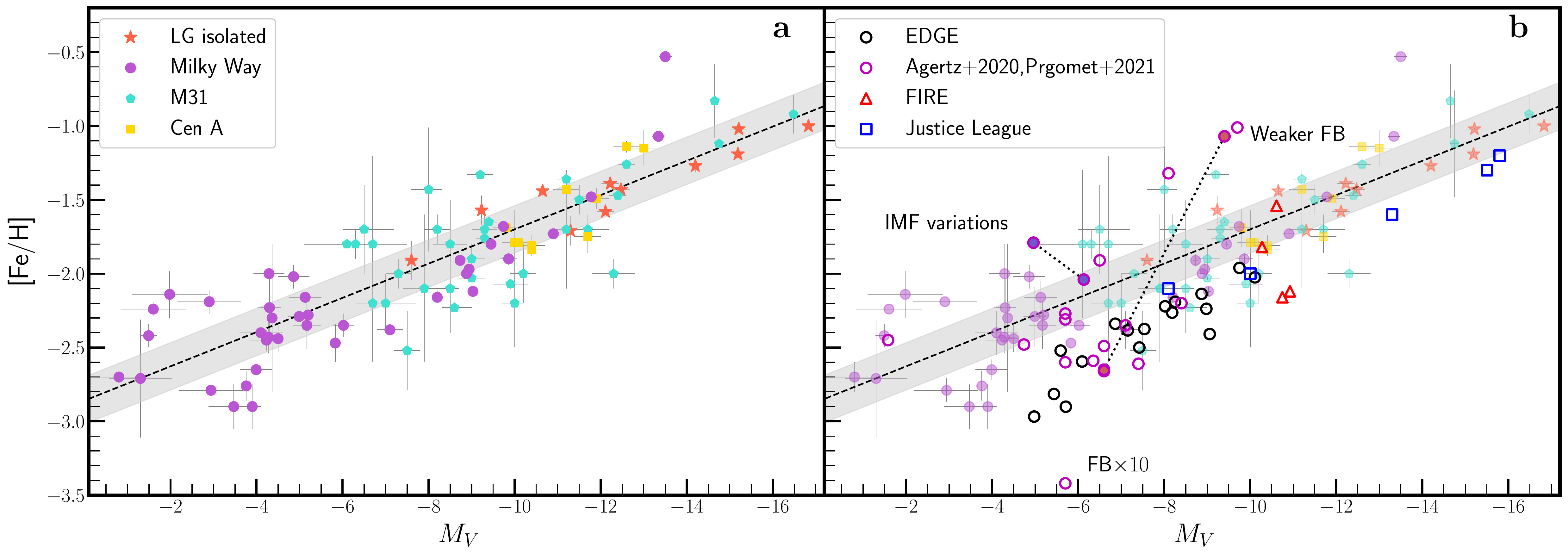}
\caption{{\bf The observed stellar [Fe/H]-$M_V$ relation for dwarf galaxies}. {\bf a:} Observations from the Milky Way dwarf galaxies (purple circles \cite{Simon19}, Kim et al. 2021 in prep.), isolated Local Group dwarfs (orange stars, \cite{geha10,kirby13,taibi18,taibi20}), M31 (cyan pentagons, \cite{collins13,collins20,collins21,wojno20}) and Centaurus A (yellow squares, \cite{mueller21}) are shown. All uncertainties in $M_V$ and [Fe/H] are taken from the works cited. The dashed line and grey shaded regions show the best-fit relation and $1\sigma$ scatter from \cite{kirby11}. {\bf b:} A comparison of same [Fe/H]-$M_V$ observations, and \cite{kirby11} relation as panel a, compared with recent numerical simulations of dwarf galaxies, as marked. The [Fe/H]-$M_V$ relation is particularly sensitive to feedback models, with stronger/weaker feedback moving galaxies $\sim$vertically in this space (dotted line connecting the brown circles), and IMF variations moving galaxies $\sim$orthogonal to the relation (dotted line connecting the blue circles).}
\label{fig:feh}
\end{figure}

Fig. \ref{fig:feh} shows the observed [Fe/H]-$M_V$ relation for the Milky Way (purple circles \cite{Simon19}, Kim et al. 2021 in prep.), isolated Local Group dwarfs (orange stars, \cite{mcconnachie12}), M31 (cyan pentagons, \cite{collins13,collins20,collins21,wojno20}) and Centaurus A (yellow squares, \cite{mueller21}). The black dashed line and grey band shows the relation and its 68\% confidence interval, respectively, that is known to extend also to higher masses \citep{kirby13}.

In the right panel of Fig. \ref{fig:feh}, we compare the observed [Fe/H]-$M_V$ relation with a host of recent state-of-the-art numerical simulations: EDGE (\cite{agertz20,orkney21,prgomet21}; Kim et al. in prep. 2021); FIRE \citep{fitts17}; and Justice League \citep{applebaum21}. \cite{agertz20} show that the [Fe/H]-$M_V$ relation is uniquely sensitive to stellar feedback physics. They run the same low-mass dwarf ($M_{200} \sim 10^9$\,M$_\odot$), dialing up and down the feedback strength (dotted line connecting the brown circles). Their fiducial simulation -- with a feedback strength consistent with current data from star forming regions -- is consistent with current data. Stronger feedback ejects more metals per unit stellar mass causing the galaxy to fall significantly below the [Fe/H]-$M_V$ relation. Conversely, weaker feedback ejects too few metals, the galaxy over-enriches and it moves above the relation. Similarly, motivated by tentative observational evidence \citep[e.g.][]{geha13}, \cite{prgomet21} explore the impact of IMF variations on the lowest mass dwarfs. A more top-heavy IMF boosts the SN feedback, but also increases the metal production due to the presence of more massive stars. The net result is higher metallicity dwarfs at lower $M_V$ (dotted line connecting the two blue circles).

Interestingly, all of the latest simulations appear to lie slightly below the [Fe/H]-$M_V$ relation. At low $M_V$, this may be tentative evidence for IMF variations \citep{prgomet21}, it could point to current feedback models being overly strong, or it could indicate that the lowest mass dwarfs are more tidally affected than their current orbits suggest. At higher $M_V$, it could point to the important role of radiative transfer (RT). \cite{agertz20} find that models with RT have slightly higher [Fe/H] at fixed $M_V$ than those without. This is because RT keeps gas warm in between starbursts, leading to less vigorous outflows and, therefore, fewer metals escaping into the IGM. Cosmic ray feedback may produce a similar effect \citep[e.g.][]{Booth13}. Alternatively, such discrepancies could owe to problems with the assumed chemical yields. \cite{applebaum21} find no such offset when computing the total metals rather than just iron. Either way, it is clear that the [Fe/H]-$M_V$ relation encodes a wealth of information about stellar feedback physics that can, at least in principle, be used to test and hone models.

\subsection{Distribution and retention of metals in dwarf galaxies}
\label{sec:gas}

The stellar metallicity is only one tracer of the total metal budget of dwarf galaxies. For gas-rich systems, we can also measure the metallicity of their ISM and (in some cases) CGM. Further, given their low-mass potentials, we expect that dwarf galaxies will lose a high fraction of their metals through outflows ($\gtrsim50\% $ \cite[e.g.][]{maclow99,muratov17}), distributing them to the CGM (and potentially the IGM). Mapping the distribution of metals from the stars to the ISM and CGM is therefore key for constraining feedback mechanisms.

For those nearby galaxies with detailed stellar metallicities, we often have no comparable gas metallicity, as the majority of local dwarf galaxies are quenched systems, devoid of gas. Estimates of the amount of metals lost can be obtained using their stellar abundances and chemical evolution modelling. \cite{kirby11} performed such a study with 8 Milky Way dwarf galaxies and estimated they had ejected $\sim96-99\%$ of their metals. Gas-phase metallicities, however, can be measured for systems much farther afield and can provide a more complete accounting of the metals in a dwarf. \cite{berg12} conducted a large scale spectroscopic survey of low-mass local volume dwarfs to measure the oxygen abundances in their H~II regions. Using a robust sample of 38 objects, they showed a tight relation between luminosity and gas-phase metallicities for these objects. Studies such as these present another strong test for feedback models. 

A few isolated Local Group dwarfs are gas-bearing, allowing us to compare the fraction of metals in their stars, ISM and CGM directly. \cite{zheng19} study the CGM of WLM -- which has a stellar mass of $M_*=4.3\times10^7\,{\rm M_\odot}$ \cite{mcconnachie12} -- and find that the majority of metals by mass likely reside there (15-77\%), with only $\sim3\%$ locked in the stars and 6\% in the ISM. This is consistent with expectations from simulations, e.g. \cite{muratov17}, who show that $\gtrsim50\%$ of metals are found in the CGM for dwarfs of this stellar mass. Another example is Leo P, which has a stellar mass of $M_*=5.6\times10^5\,{\rm M_\odot}$ \cite{mcquinn15c}. \cite{mcquinn15} measure the gas phase oxygen abundance for Leo P and determine how much oxygen it should have produced based on its resolved SFH. They find that Leo P has retained only 5\% of the oxygen it has produced, 25\% of that is in stars, while 75\% is in the gas phase. This is compared to the 25\% retained in more massive galaxies(\cite[e.g.][]{peeples14}.

\section{The future of the field}\label{sec:future}

Throughout this review, we have identified key observables with which to confront theoretical models. In this final section, we discuss how future facilities will allow us to make significant progress in mapping stellar feedback and its impact on dwarf galaxies. 

As highlighted in \S\ref{sec:mfeh}, simulations still tend to under-predict the metallicities of the faintest dwarfs. Understanding this offset is a key challenge. To rule out tidal effects skewing the observational trend, a large sample of isolated ultra-faint dwarfs with measured chemistries is required. The 8.4mVera Rubin Observatory will  shortly begin its 10 year Large Survey of Space and Time (LSST), optically mapping the southern skies. This survey will allow for the detection of ultra-faint dwarfs out to distances exceeding 1\,Mpc \cite{tollerud08}, where they will be unaffected by ram pressure and tidal stripping from large host galaxies such as the Milky Way. Depending on their distances, these objects can be followed up with existing 10-m or future 30-m telescopes such as the European Extremely Large Telescope. If the offset remains, it may indicate that IMF variations are important for modelling the metallicities of the faintest dwarfs.

Secondly, do dwarf galaxies quenched solely by SN exist? A volume complete sample of isolated dwarf galaxies from the Vera Rubin observatory will allow us to hunt for these. To determine exactly when they quenched, precision SFH are needed (using either HST or future space telescopes). However, age resolution at the furthest lookback times is typically coarse even for those nearby galaxies with deep imaging. Progress is rapidly being made in the stellar evolutionary models underpinning these studies \cite[e.g.][]{gallart21}. Further progress on this will help definitively test the impact of external and internal feedback on the faintest galaxies.

Thirdly, new mass modelling techniques are needed that can handle the complications that feedback introduces (e.g. HI holes) when measuring rotation curves and dark matter masses if we hope to fully exploit the wealth of HI data that will soon come from the SKA/WALLABY. These surveys will give a large, volume complete survey of faint, HI-rich dwarf galaxies with a known selection function. But without robust tools, we cannot fully exploit this exciting observational resource.

Finally, a number of upcoming surveys and facilities will allow wide-scale multi-band mapping of outflows and metallcities of dwarf galaxies. The current HST ULYSSES \cite{romanduval20} and CLASSY \cite{berghst} programs are capturing UV spectra of high-mass stars and low-metallicity galaxies, allowing the study of their winds and outflows. The SPRITE UV cubesat is due for launch in 2022, and will map ionizing radiation escaping from star-forming galaxies \cite{fleming19}. Future integral-field spectrographs installed on ground-based observatories will allow the tracking of galactic winds and outflows at sensitivities not previously permitted. These include the proposed BlueMUSE spectrograph for the Very Large Telescope \cite{richard19}. It's high-resolution and sensitivity to the blue and UV will make it a powerful probe of feedback and its impact on the ISM and CGM in star-bursting dwarfs. In the 2040s, the US expect to launch the large IR/O/UV telescope (recommended in the US 2020 Decadal survey). This will allow multi-band studies of the stellar and gas contents of dwarf galaxies to large distances, placing further constraints on outflows, mass loading factors and metal distributions over a wide range of redshifts and environments. Such observations will permit a test of feedback predictions from cosmological simulations which currently seem to over-estimate mass loading in the lowest mass dwarfs (as discussed in \S~\ref{sec:massload}) and under-estimate their starburst duration (as discussed in \S~\ref{sec:sfh}).

The expected advances in instrumentation and models over the next decade listed above should allow significant progress in understanding the properties of, and mechanisms powering  galactic winds in dwarf galaxies. They will also allow a measure of the impact they have on the structure, chemistry and dark matter content of these faint galaxies, and lead to a wider understanding of the impact of reionisation on low mass galaxies, and the movement of metals into the CGM. \\

\noindent Correspondence and requests for materials should be addressed to Michelle Collins (m.collins$@$surrey.ac.uk).

\section*{Data Availability}

All the data that support the findings of this study are taken from published works (referenced in the text) and are collated into a GitHub repository, \href{https://github.com/justinread/feedback_perspective}{https://github.com/justinread/feedback\_perspective}.

\section*{Code Availability}
The code needed to reproduce all figures in this perspective can be found in the same GitHub repository as the data above, \href{https://github.com/justinread/feedback_perspective}{https://github.com/justinread/feedback\_perspective}

\bibliography{feedback}

\section*{Acknowledgements}
We would like to thank the 4 anonymous reviewers for helpful and insightful comments which have improved this manuscript. 

\section*{Author contributions}
The authors contributed equally to all aspects of the article.

\section*{Competing interests}
The authors declare no competing interests. 

\section*{Publisher’s note}
Springer Nature remains neutral with regard to jurisdictional claims in published maps and institutional affiliations.

\section*{Supplementary information (optional)}
If your article requires supplementary information, please include these files for peer-review. Please note that supplementary information will not be edited.

\newpage

\end{document}